# Strain-tunable van der Waals interactions in few-layer black phosphorus

Shenyang Huang[1,2], Guowei Zhang[1,2], Fengren Fan[1,3], Chaoyu Song[1,2], Fanjie Wang[1,2], Qiaoxia Xing[1,2], Chong Wang[1,2], Hua Wu[1,3] & Hugen Yan[1,2]*.

[1]State Key Laboratory of Surface Physics and Department of Physics, Fudan University, Shanghai 200433, China.

[2]Key Laboratory of Micro and Nano -Photonic Structures (Ministry of Education), Fudan University, Shanghai 200433, China

[3]Key Laboratory for Computational Physical Sciences (Ministry of Education), Fudan University, Shanghai 200433, China.

*E-mail: hgyan@fudan.edu.cn

**Abstract**

Interlayer interactions in 2D materials, also known as van der Waals (vdWs) interactions, play a critical role in the physical properties of layered materials. It is fascinating to manipulate the vdWs interaction, and hence to "redefine" the material properties. Here, we demonstrate that in-plane biaxial strain can effectively tune the vdWs interaction of few-layer black phosphorus with thickness of 2-10 layers, using infrared spectroscopy. Surprisingly, our results reveal that in-plane tensile strain efficiently weakens the interlayer coupling, even though the sample shrinks in the vertical direction due to the Poisson effect, in sharp contrast to one's intuition. Moreover, density functional theory (DFT) calculations further confirm our observations and indicate a dominant role of the puckered lattice structure. Our study highlights the important role played by vdWs interactions in 2D materials during external physical perturbations.

**Introduction**

Recently, layered materials such as graphene, transition metal dichalcogenides (TMDCs) and black phosphorus (BP) have attracted increasing attention. The layers in these materials are held together by van der Waals (vdWs) interactions, which play a critical role in the material property, and hence device performance. One of the most fascinating aspects is that different two-dimensional (2D) layers can stick together through weak vdWs interactions without the constraint of lattice match, holding great promise for building hetero-structure-type materials with entirely new properties[1,2]. A variety of novel quantum phenomena have been reported in vdWs heterostructures, such as fractional quantum Hall effect[3,4], gap opening in gapless graphene[5], unconventional superconductivity[6], tunable Mott insulator[7,8], long life time Moire excitons[9] and so on. Therefore, it provides us a new avenue to tailor the properties of layered materials by manipulating vdWs interactions. The strength of vdWs interaction is closely related to the interlayer distance, which can be shrunk or expanded by external perturbations, such as hydrostatic pressure[10,11] and ion intercalation[12,13]. Only recently, researches showed that in-plane strain can couple with interlayer interactions in bilayer $MoS_2$[14] and $MoS_2/WS_2$ heterostructure[15] through Raman and photoluminescence spectroscopies.

BP is an emerging layered material with puckered lattice structure and thickness dependent direct bandgap[16-20]. The van der Waals interaction in few-layer BP splits a band into multiple subbands and causes a series of optical resonances in the infrared (IR) spectra[17,19] which provides a unique opportunity to monitor the interlayer coupling strength. Experiments have shown that few-layer BP is very sensitive to in-plane strain, with a large strain tunability in the band gap and higher energy optical transitions[17,21,22]. However, there is no study on the strain effect on the interlayer coupling up to date. Typically, in-plane strain tunes the in-plane bond strength and how it affects the interlayer coupling is not clear. In this paper, we systematically investigate the evolution of band structures of few-layer BP under in-plane biaxial strain, using IR absorption spectroscopy. We separate the interlayer and intralayer contributions to the evolution of the bandstructure under in-plane strain and gain quantitative insights into the van der Waals coupling effect in the strain engineering process. Surprisingly, we found that in-plane straining is a very efficient scheme to tune the vdWs interaction in BP, which opens a promising avenue for strain engineering of vdWs interactions.

**Results**

**Sample characterization and the scheme for biaxial strain**

We studied exfoliated BP samples on polypropylene (PP) substrate (see Methods for sample preparation). The IR extinction ($1-T/T_0$) spectrum of few-layer BP was obtained using Fourier transform infrared spectrometer, where $T/T_0$ is the ratio of light transmission with/without BP flakes (see details in Methods)[17]. From the polarized IR spectrum, we can directly identify the layer number and crystal orientation[17]. Figure 1b is a typical IR extinction spectrum of a 6L BP. Two prominent peaks can be observed, labeled as $E_{11}$ (transition from the first valence band $v_1$ to the first conduction band $c_1$) and $E_{22}$ transitions ($v_2$ to $c_2$), as indicated in the insert of Fig. 1b. Biaxial tensile (compressive) strain was introduced to the BP sample through heating (cooling) and strain value was determined according to the experimentally measured thermal expansion coefficient of PP substrate (see Supplementary Figs. 1 and 2 for details). To avoid possible slippage between BP flakes and PP substrates as well as non-uniform deformation of PP, the temperature variation was kept within a small range of 25-75℃ (corresponding to the biaxial strain range of -0.3 - 0.3%, see Supplementary Fig. 2 for the reason of compressive strain).

**Transition index-dependent strain effect**

Figure 1c shows the typical IR extinction spectra of a 6L BP sample under various biaxial tensile and compressive strains. Both of the characteristic peaks ($E_{11}$ and $E_{22}$) exhibit blueshift with tensile strain and redshift with compressive strain. Interestingly, the $E_{11}$ peak shifts faster than the $E_{22}$ peak, with shift rates of 222 and 167 meV/%, respectively, as extracted from linear fittings in Fig. 1d. The insert of Fig. 1d shows that the energy difference between $E_{11}$ and $E_{22}$ decreases as tensile strain is applied. Although optical resonances are dominated by excitons in low dimensional semiconductors[19,20], the shift is mainly attributed to the change of the electronic band structure and the modification of exciton binding energy due to such small strain is negelected[23]. In addition, the observed shift of resonances might be a mixed result of strain and pure temperature effect. To account for this possibility, controlled experiments were carried out. We transferred few-layer BP samples to quartz substrates, which have a much smaller thermal expansion coefficient than PP. Therefore, the strain effect of quartz can be ignored. As shown in Supplementary Fig. 3, the pure temperature effect in the range of 25-75 ℃ is almost negligible compared to the strain effect. Thus, in our analysis, we simply neglect the pure temperature-induced shift. The fact that $E_{11}$ shifts faster than $E_{22}$ is not limited to biaxially strained 6L and can be observed in all strained few-layer BP samples, even uniaxially strained samples (See more data in Supplementary Figs. 4 and 5). Generally, we find that optical resonances originated from higher-index subbands exhibit smaller shift rate under strain. This transition index-dependent strain effect indicates that interlayer vdWs interaction changes with in-plane strain, since the splitting of subbands directly reflects the strength of interlayer coupling. Quantitative description will be provided in the Discussion part.

**Layer-dependent strain effect**

For $E_{11}$ transitions of samples with different thickness, the shift rates are different as well. We can obtain BP samples with different thickness in one big flake, which are presumably under the same strain condition all the time. An example is shown in Fig. 2a, where 3L and 4L BP are adjacent to each other. To confirm this, Raman measurements were performed. According to previous studies[24], it's reasonable to assume that the strain effect on some of the Raman peaks has little sample thickness dependence. Therefore, if the shift rates of Raman peaks under strain are the same, we can guarantee the same strain magnitude for different BP layers all the time. Given that there is Davydov splitting of $A_g^2$ mode[24,25], and the intensity of $B_{2g}$ mode is relatively weak under excitation with light polarized along armchair direction, $A_g^1$ peak shift serves as an indicator for the strain magnitude.

Indeed, the shift rates of the $A_g^1$ mode are almost the same, with values of -5.17

and -4.92 cm$^{-1}$/% for 3L and 4L, respectively (see Supplementary Fig. 8), which assures the same strain condition. Figure 2b shows the IR extinction spectra of this sample under different strains, with peak positions as a function of biaxial strain summarized in Fig. 2c. $E_{11}$ of 3L shifts more slowly than that of 4L, with shift rates of 158 and 185 meV/%, respectively. This result shows that the shift rate of $E_{11}$ transition also depends on the sample thickness. Indeed, this dependence is systematic, as evidenced by a large number of measured samples with thickness ranging from 2-10L. The averaged shift rates for each layer number are plotted in Fig. 3a, with the error bar indicating the shift rate range of multiple measured samples (for each layer number, at least three samples were measured to ensure the reproducibility of the data). To make sure the strain was accurately calibrated for different thickness samples, the Raman $A_g^1$ peak shift induced by strain was served as a monitor. As we can see in Fig. 3a, the shift rates of $E_{nn}$ transitions are strongly subband index and layer number dependent, particularly for thin samples (layer number below 6L). Our measurements reveal that thinner samples and transitions associated with higher-index subbands are less sensitive to strain (smaller shift rates), which implies that in-plane strain can tune the vdWs interaction, as detailed in the Discussion part.

To further confirm this, we performed uniaxial strain experiments for BP with different layer number, with the strain direction along AC and ZZ, respectively. Results are presented in Supplementary Figs. 6 and 7. Notable layer number dependence of the strain effect is observed both for AC and ZZ strains, consistent with biaxial strain results. Note that the shift rates induced by uniaxial strain are almost half of those induced by biaxial strain, which is as expected since the sample deforms in two directions under biaxial strain while only one direction under uniaxial strain. According to Fig. 3a, we can infer that the shift rate of bulk samples under biaxial strain is a little larger than 200 meV. This suggests a shift rate of ~100 meV/% under uniaxial strain, which is in good agreement with previous results (99 meV/% for strain along AC and 109 meV/% for strain along ZZ in Ref. 21).

**Discussion**

For unstrained few-layer BP, a 1D tight binding model accounting for the interlayer coupling can well describe the optical transitions, with only two relevant parameters: the monolayer band gap $E_g$ and the interlayer coupling strength[17,19]. It's apparent that $E_g$ depends on the in-plane strain, since it's merely determined by the intralayer bond. The dependence of interlayer van der Waals interaction on the in-plane strain is not clear yet. An intuition may imply that the interlayer coupling would be enhanced with in-plane tensile strain since the interlayer distance typically decreases due to the Poisson effect. However, our experiment gives opposite result, as detailed in the following analysis. From 1D tight binding model (see details in Supplementary Information), the shift rate of transition energy due to strain is

$$\frac{dE_{nn}}{d\varepsilon} = h - 2k\cos(\frac{n\pi}{N+1}), \quad (1)$$

where $\varepsilon$ is the applied biaxial strain, $N$ is the layer number, $n$ is the index of subband, $h$ and $k$ are fitting parameters describing the change rates of monolayer band-gap and van der Waals interaction strength under strain, respectively. Based on this equation, we can conclude that the shift rate of transition energy is layer (*N*) dependent and transition index (*n*) dependent for $k \neq 0$.

To extract the values of *h* and *k*, we used equation (1) to globally fit all of the experiment data, shown as the solid curves in Fig. 3a. The fitting values of *h* and *k* are 66 and -86 meV/%, respectively. The overall trend of the fitting curves is in reasonable agreement with the experiment data. The deviation of data points from the fitting curves is mainly attributed to the experimental uncertainty in determining the strain value. Of course, equation (1) doesn't take into account excitonic effects and other many-body interactions, which may compromise its accuracy. Nevertheless, the basic behavior is well captured by the model. Since *h* and *k* are in the same order of magnitude, the modification of van der Waals interaction under in-plane strain contributes as significantly as that of intralyer bonding does to the tuning of the band structure. Previously, we experimentally determined the transition energy $E_{nn}=E_{g0} - 2\Delta\gamma\cos(n\pi/(N+1))$, with $\Delta\gamma$=880meV, for unstrained few-layer BP[17]. We can see that 1% in-plane strain changes the interlayer coupling ($\Delta\gamma$) almost 10%, which is quite remarkable.

To further confirm our experimental results, we performed DFT calculations of the strain effect. The details of the calculation are presented in the Methods. Figure 3b shows the DFT calculated shift rates for 1L, 2L, 3L and 8L BP, in good agreement with our experiments (see Fig. 3a). The infinite lattice periodicity in the *ab* plane (in-plane) gives rise to dispersive bands, which have the band gap at $\Gamma$ point. For those few-layer BPs, the interlayer vdWs interaction along the *c*-axis (out-of-plane) lifts the energy degeneracy at $\Gamma$, thus forming different groups of subbands. For example, we plot in Fig. 3c the lattice structure of a 2L BP, and in Fig. 3d the illustrative band structure of 2L BP and its evolution against strain. When an in-plane tensile strain is applied (see the right part of Fig. 3d), the increasing intralayer atomic distance and thus the decreasing intralayer interaction reduces the band width and enlarges the band gap. Moreover, our calculation also indicates a decrease of the interlayer interaction, since the subband splitting gets smaller. As a result, for the lower-index subband transitions ($E_{nn}$, $n<N/2$) in the finite $N$-layer BP, both the decreasing intralayer and interlayer interactions contribute positively to the increasing $E_{nn}$, giving a positive energy shift rate against the strain. However, for the higher-index transitions ($E_{nn}$, $n>N/2$), the decreasing intralayer (interlayer) interaction yields a positive (negative) contribution to the $E_{nn}$, making its shift rate smaller than the lower-index $E_{nn}$. All these mechanisms are well captured by the DFT calculations: for a given few-layer BP, the energy shift rate of $E_{nn}$ gets smaller with the increasing index, and $E_{nn}$ can even become negative, e.g., $E_{33}$ for 3L BP and $E_{nn}$ ($n \geqslant 5$) for 8L BP. There is overall a quite good agreement between the DFT calculations and the experiments, as shown in Figs. 3a and 3b.

It should be emphasized that the fitting parameter $k$ in Fig. 3a is negative. A negative $k$ means a weakened interlayer coupling under tensile strain. This is rather astonishing, given that in-plane tensile strain typically compresses the sample in the out-of-plane direction due to the Poisson effect. Now we see that a shorter distance between layers gives a weaker van der Waals coupling between them. How could this happen? In order to make sense out of it, we have to revisit the puckered lattice structure of few-layer BP.

According to the tight-binding model, the band structure in the vicinity of $\Gamma$ point of the Brillouin zone in few-layer BP is mainly determined by three hopping parameters: two intralayer ones ($t_{\parallel}^1$ and $t_{\parallel}^2$) and one interlayer one ($t_{\perp}$)[26], as illustrated in Fig. 3c. The hopping parameter $t$ is generally related to the bond length $r$ in the form of $t \propto \frac{1}{r^2}$ (Ref.[27]). It should be noted that the vdWs interaction between layers is a complicated and sometimes vague concept. Here we just capture the major contributing factor $t_{\perp}$ for our discussion, and treat it on the same footing as in-plane bonds. For monolayer BP, the bandgap can be expressed as $E_g = 4 t_{\parallel}^1 + 2 t_{\parallel}^2$. While for 2L (or few-layer) BP, interlayer interaction ($t_{\perp}$) also contributes to the band structure, and leads to the splitting of valence and conduction bands, resulting in two subband transitions: $E_{11}$ and $E_{22}$. The energy difference between $E_{11}$ and $E_{22}$ ($\Delta = E_{22} - E_{11}$) is proportional to $t_{\perp}$ (Ref.[26]).

Based on the above analysis, one can directly understand how the band structure of 2L BP evolves with the in-plane biaxial strain. Under biaxial tensile strain, the absolute values of $t_{\parallel}^1$ and $t_{\parallel}^2$ both decrease. But it is still not straight forward to predict how $E_g$ changes since the two hopping parameters have different signs ($t_{\parallel}^1 < 0$, $t_{\parallel}^2 > 0$) (Ref.[26]). However, the experiment results show that both $E_{11}$ and $E_{22}$ blueshift, indicating biaxial tensile strain induces an increase of $E_g$. Moreover, when tensile strain is applied, atoms connected by $t_{\perp}$ are split apart laterally, since the two atoms are not in registry (one right above the other). More importantly, those two atoms are split apart vertically as well, even though the Poisson effect compresses the sample in the vertical direction, as shown by our DFT calculations (see Supplementary Fig. 9).

This unusual scenario originates from the puckered structure within each BP layer, which is constituted of two sub-layers. When biaxial in-plane tensile stain is applied, the average distance between two adjacent layers ($D+d$, shown in Fig. 3c) decreases due to the Poisson effect. According to the DFT calculations (Supplementary Fig. 9), when 1% biaxial tensile strain is applied, the interlayer distance ($D+d$) between two phosphorene layers decreases by 0.031 Å, corresponding to an out-of-plane Poisson's ratio $\nu = -\frac{d\varepsilon_z}{d\varepsilon_{xy}}$ ~0.5. However, at the same time, the distance between two sub-layers ($d$) within each layer decreases even more thanks to the puckered structure (0.087 Å, with 1% biaxial tensile strain, see Supplementary Fig. 9), causing an increased gap ($D$) between two layers (0.056 Å, with 1% biaxial tensile strain). As a result, the distance between the two atoms responsible for $t_\perp$ increases, leading to the decrease of $t_\perp$ and hence the decrease of interlayer interactions. The puckered structure of BP efficiently facilitates this counterintuitive phenomenon. Therefore, when the interlayer interaction decreases, the energy difference between $E_{11}$ and $E_{22}$ decreases. Meanwhile, tensile strain leads to the blueshift of both $E_{11}$ and $E_{22}$, hence the shift rate of $E_{11}$ is larger than that of $E_{22}$, as illustrated in Fig. 3d. With above argument, we conclude that the higher the transition index is, the smaller the shift rate is. As demonstrated by DFT calculation in Fig. 3b, the shift rate even changes to negative sign when the subband index $n>5$ for the 8L BP. The split subbands of few-lay BP provide us a unique platform to monitor the van der Waals interactions under strain or other physical perturbations.

In summary, we systematically investigate the evolution of band structures in few-layer BP under in-plane biaxial strain. Our results show that strain effect of few-layer BP exhibits layer number and subband index dependence. This dependence is closely related to the tuning of interlayer coupling. Surprisingly, in-plane biaxial tensile strain weakens such coupling, which is counterintuitive. Indeed, it is the puckered structure of BP which facilitates this phenomenon. Our study paves the way for strain engineering of vdWs interactions in few-layer BP and other 2D materials.

**Methods**

**Sample preparation and the strain setup**

Few-layer BP flakes were mechanically exfoliated from bulk crystals (HQ graphene Inc.) onto PDMS substrate, and then transferred to PP substrates by a dry-transfer method. To apply biaxial strain, the sample was loaded into a heating and cooling chamber (Linkam, FTIR600). To avoid sample degradation, all sample preparation procedures were conducted in a glove box filled with $N_2$ gas ($O_2$ and $H_2O$ < 1 ppm). For optical measurements, which were outside of the glove box, the chamber was flushed with $N_2$ gas to protect BP samples.

**Optical measurements**

The IR extinction (1-$T/T_0$) spectrum of few-layer BP was recorded using a Fourier transform infrared spectrometer (FTIR, Bruker Vertex 70V) in conjunction with an infrared microscope (Hyperion 2000). By using tungsten halogen lamp as IR source, quartz beam splitter and liquid nitrogen cooled MCT detector, wide range extinction spectrum from 0.25 to 1.36 eV could be obtained. The incident light was polarized along the armchair (AC) direction through a broadband ZnSe grid polarizer. The visible reflection ($\Delta R/R_0$) measurements of few-layer BP were performed using an Andor grating spectrometer (SR550i) in conjunction with a Nikon inverted microscope (Elipse Ti-U). By using a broadband LED light source and a CCD detector, the visible reflection spectrum ranging from 1.55 to 3.14 eV could be covered. Raman measurements were conducted using a commercial Raman system (Horriba, Jobin Yvon HR-Evolution 2). 532 nm and 633 nm lasers were available for excitation.

**DFT calculations**

All the DFT calculations were performed using the Vienna ab initio simulation package (VASP)[28,29]. The generalized gradient approximation of Perdew-Burke-Ernzerhof (PBE)[30] type was employed for the exchange-correlation energy. The interlayer interactions, Van der Waals interactions, were included by using the optB88 vdW density functional[31,32]. To describe the band gaps more accurately, the Heyd-Scuseria-Ernzerhof (HSE) hybrid functional[33] was also used. Wave functions were expanded in terms of plane waves, for which the kinetic energy cutoff was 500 eV, and the energy integration was performed on a $10 \times 12 \times 1$ Monkhorst-Pack k-point grid. All the structures in our calculations were fully optimized with a force criteria of 0.01 eV/Å. For the calculations of all the few-layer BPs, slab models were employed, in which a 20 Å vacuum was added along the $c$ direction.

# Data availability

The data that support the findings of this study are available from the corresponding author upon request.

## Acknowledgments

H.Y. is grateful to the financial support from the National Young 1000 Talents Plan, National Science Foundation of China (Grant Nos. 11874009, 11734007), the National Key Research and Development Program of China (Grant Nos. 2016YFA0203900 and 2017YFA0303504), Strategic Priority Research Program of Chinese Academy of Sciences (XDB30000000), and the Oriental Scholar Program from Shanghai Municipal Education Commission. H. W. is supported by the National Science Foundation of China (Grants No. 11474059 and No. 11674064) and the National Key Research and Development Program of China (Grant No. 2016YFA0300700). G.Z. acknowledges the financial support from the National Natural Science Foundation of China (Grant No. 11804398) and Open Research Fund of State Key Laboratory of Surface Physics. C.W.

is grateful to the financial support from the National Natural Science Foundation of China (Grant No. 11704075) and China Postdoctoral Science Foundation. Part of the experimental work was carried out in Fudan Nanofabrication Lab.

## Author contributions

H.Y., G.Z. and S. H. initiated the project and conceived the experiments. S.H. and G.Z. prepared the samples, performed the measurements and data analysis with assistance from C.S., F.W., Q.X. and C.W. F.F. carried out DFT calculations. S. H., G.Z. and H. Y. co-wrote the manuscript with inputs from H.W. H.Y. supervised the whole project. All authors commented on the manuscript.

## Additional information

Correspondence and requests for materials should be addressed to H.Y. (hgyan@fudan.edu.cn).

## Competing financial interests

The authors declare no competing interests.

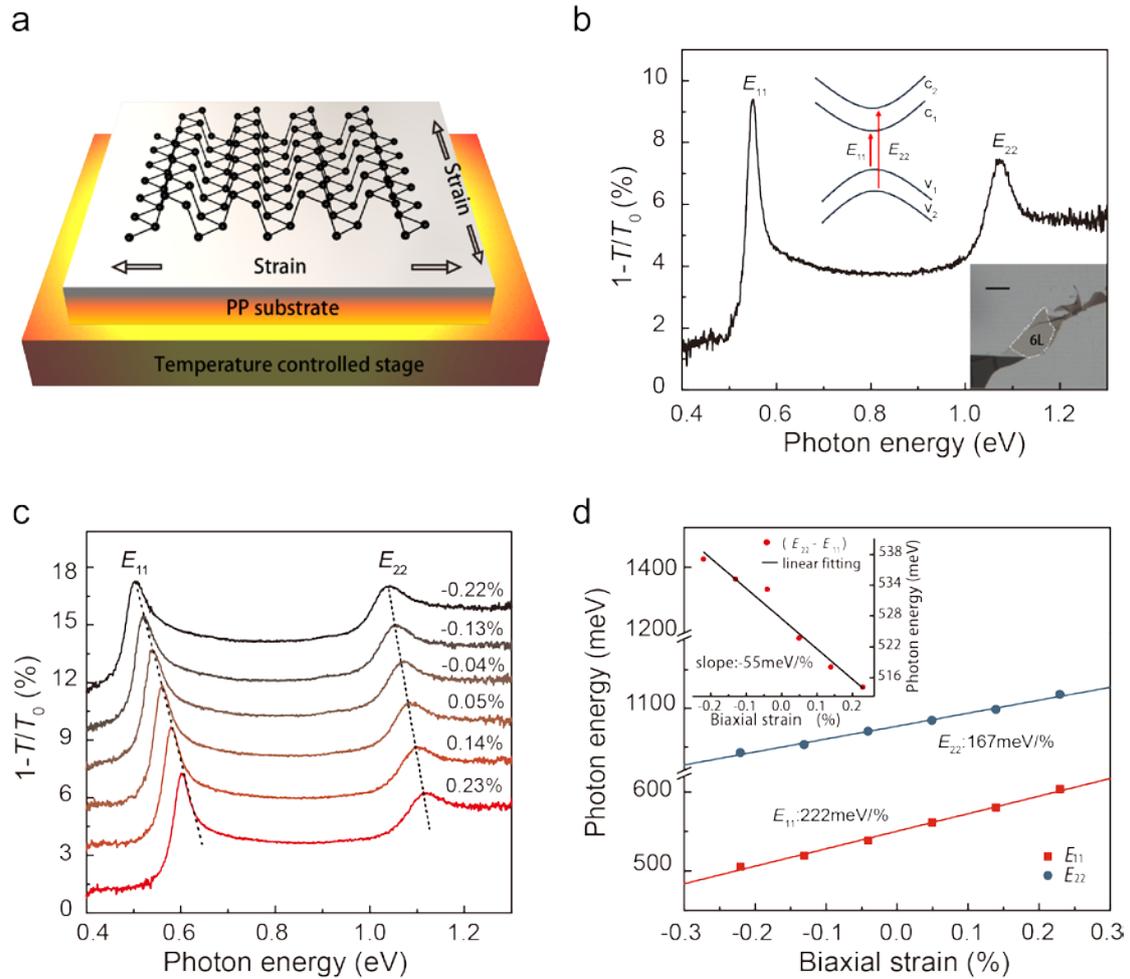

**Fig. 1 | Band structure engineering by biaxial strain** (a) Schematic illustration of the experiment setup used for applying in-plane biaxial strain by heating or cooling the PP substrate. (b) A typical IR extinction spectrum (1-$T/T_0$) for a 6L BP on PP substrate under zero strain, with the incident light polarized along the AC direction. Top insert is a schematic illustration for optical transitions between different subbands. Bottom insert is an optical image of this 6L BP sample. Scale bar: 20 μm. (c) IR extinction spectra (1-$T/T_0$) for the 6L BP under varying biaxial tensile (>0) and compressive (<0) strains. For clarity, the spectra are vertically offset. Dashed lines are guides to the eye. (d) The $E_{11}$ and $E_{22}$ peak energies as a function of biaxial strain. The solid lines are linear fits to the data which give shift rates for $E_{11}$ and $E_{22}$ of 222 and 167 meV/%,

respectively. The insert shows ($E_{22}$-$E_{11}$) as a function of biaxial strain and the solid line is a linear fit to the data.

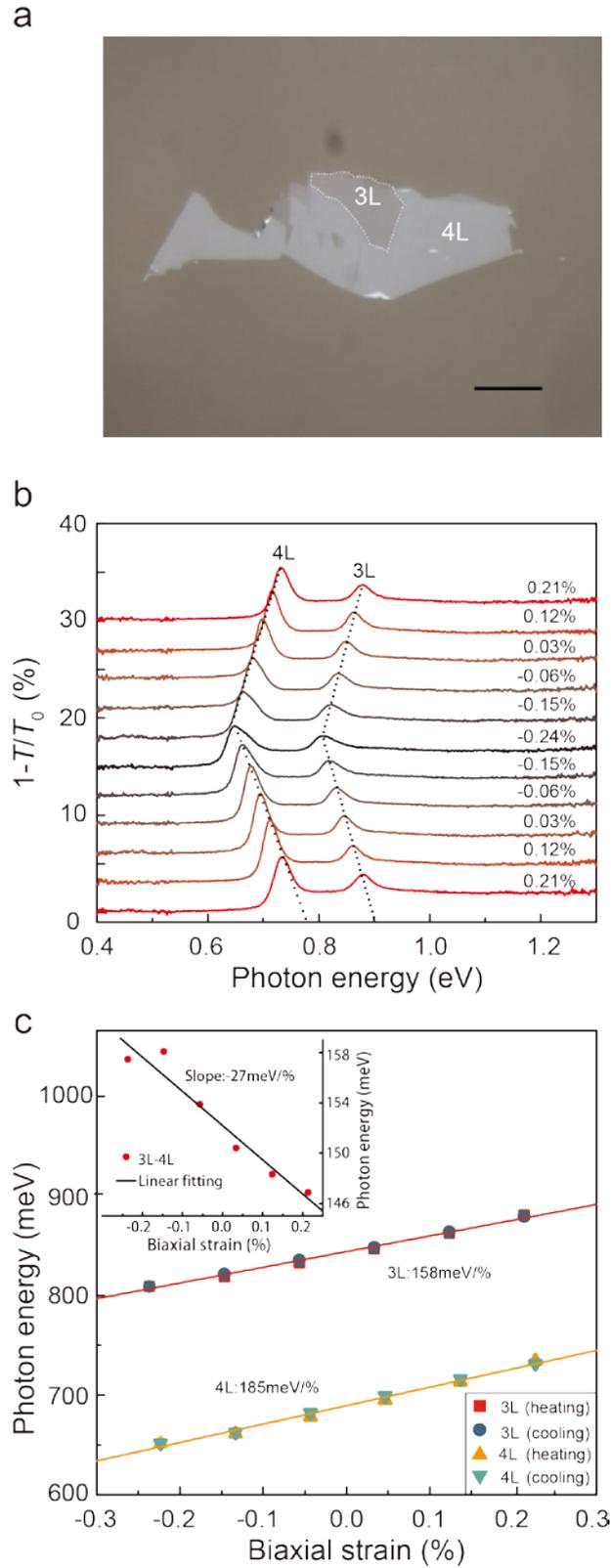

**Fig. 2 | Comparison of the biaxial strain effects on 3L and 4L BP.** (a) Optical image of the adjacent 3L and 4L BP flakes. Scale bar: 20 μm. (b) IR extinction spectra for the

3L and 4L BP under different biaxial strains. Dashed lines are guides to the eye. (c) Transition energies of $E_{11}$ in the 3L and 4L BP as a function of biaxial strain. The solid lines are linear fits to the data. The insert shows $E_{11}$ energy difference of 3L and 4L BP as a function of biaxial strain. Dots are experiment data averaged from heating and cooling processes and the solid lines are linear fits which give shift rates of 158 and 185 meV/% for 3L and 4L, respectively.

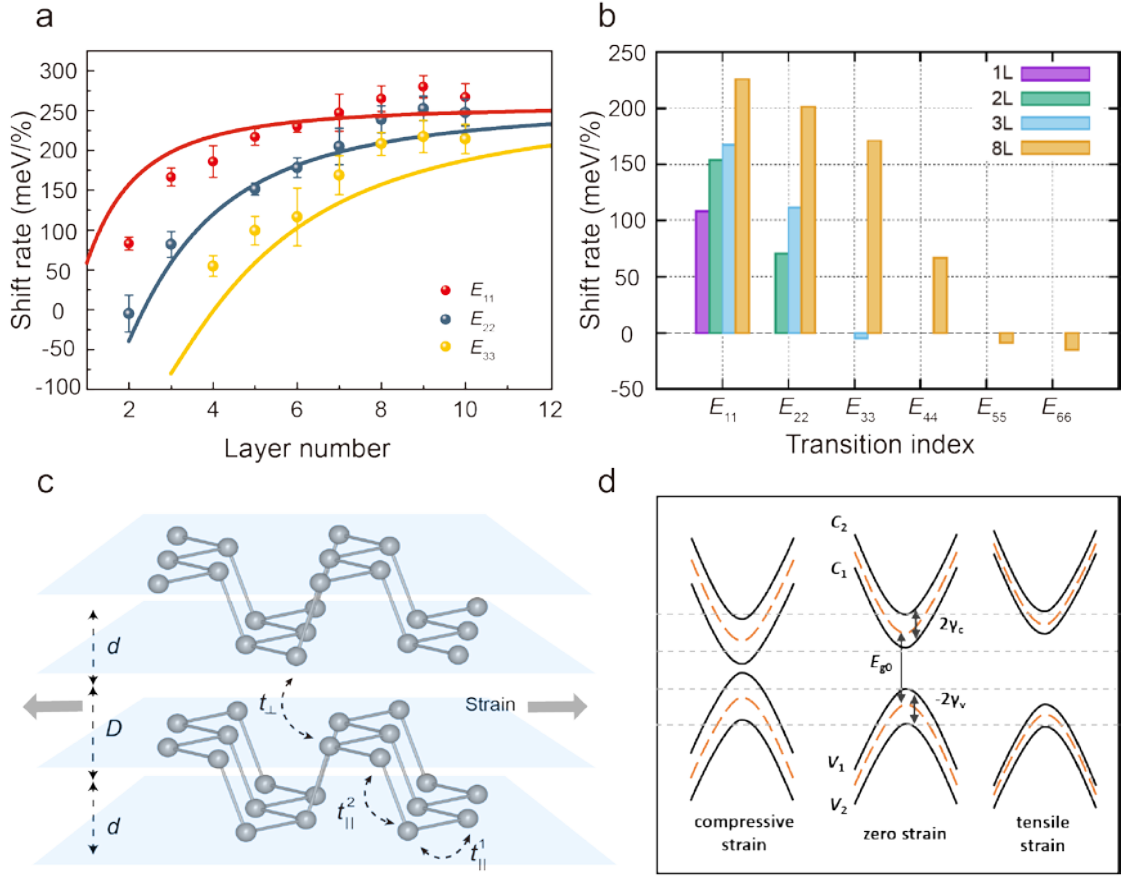

**Fig. 3 | Layer and subband index dependence of the biaxial strain effect.** (a) Averaged shift rates of $E_{11}$, $E_{22}$ and $E_{33}$ peaks as a function of layer number in 2-10L BP. The solid curves are fitted to the data using the tight-binding model shown in the text. The error bar is defined from the data spread of multiple samples. For each layer, at least three samples were measured. (b) DFT calculated shift rates for 1, 2, 3 and 8L BP induced by in-plane biaxial strain. (c) Illustration of two in-plane hoping parameters ($t_{\parallel}^{1}$ and $t_{\parallel}^{2}$) and one out-of-plane hopping parameter ($t_{\perp}$) in a 2L BP. $d$ is the height of an individual layer and $D$ is the gap between two layers. When biaxial in-plane tensile stain is applied, the average distance between two layers ($D+d$) decreases due to Poisson effect, while the gap ($D$) increases accompanied by a stronger decrease of d (see Supplementary Fig. 9). (d) Schematic illustration of the band structure evolution of a bilayer BP under tensile and compressive strain. The orange dashed curves are the bands for a monolayer BP. The change of subband splittings causes the shift rate of $E_{22}$ smaller than that of $E_{11}$.

23